# Resonant slow extraction in synchrotrons by using anti-symmetric sextupole fields

Ye Zou[1,4], Jingyu Tang[2,3,4,1], Jianquan Yang[4]

[1]University of Science and Technology of China, Hefei, Anhui 230029, PR China

[2]China Spallation Neutron Source, Institute of High Energy Physics, CAS, Dongguan 523803, China

[3]Dongguan Institute of Neutron Science, Dongguan 523808, China

[4]Key Laboratory of Particle Acceleration Physics and Technology, Institute of High Energy Physics, CAS, Beijing 100049, China

**ABSTRACT**

This paper proposes a novel method for resonant slow extraction in synchrotrons by using special anti-symmetric sextupole fields, which can be produced by a special magnet structure. The method has the potential in applications demanding for very stable slow extraction from synchrotrons. Our studies show that the slow extraction at the half-integer resonance by using anti-symmetric sextupole field has some advantages compared to the normal sextupole field, and the latter is widely used in the slow extraction method. One of them is that it can work at a more distant tune from the resonance, so that it can reduce significantly the intensity variation of the extracted beam which is mainly caused by the ripples of magnet power supplies. The studies by both the Hamiltonian theory and numerical simulations show that the stable region at the proximity of the half-integer resonance by anti-symmetric sextupole field is much smaller and flatter than the one by standard sextupole field at the third-order resonance. The particles outside the region will be driven out in two possible directions in quite short transit time but with spiral steps similar as in the third-order resonant extraction. By gradually increasing the field strength, the beam can be extracted with intensity more homogeneous than by the usual third-order resonant method, in the means of both smaller intensity variation and spike in the beginning spill. Similar to the case with a normal sextupole, we derive an empirical formula for the area of the stable region with an anti-symmetric sextupole. One can find that with the same field strength and the same tune distance to the resonance, the area of stable region or the change of the area due to the working point variation in the case of anti-symmetric sextupole is about 1/14 of the one in the case of standard sextupole. The detailed studies including beam dynamic behaviors at the proximity of other resonances, influence of 2-D field error, half-integer stop-band, and resonant slow extraction by using quadrupole field have also been presented.

## 1. Introduction

Slow extraction in synchrotrons is the main extraction method for external target experiments in particle physics and nuclear physics, and for proton or heavy ion therapy, since it could provide relatively stable beams in long time durations. Third-order resonant slow extraction [1-4] is usually used, and the principle is that one can intentionally excite the third-order resonance by controlling the tune distance (the distance between the working point and the resonance line) and the sextupole strength to peel off gradually the particles from outer to inner in the beam emittance. This method could produce relatively stable beam intensity with quite high extraction efficiency, if it is applied properly. However, this method also has some disadvantages, e.g. the tune should be moved very close to the resonance line before extraction and the stable region of the beam in the phase plane when the resonance is excited is very sensitive to the tune's stability. If we want to obtain highly stable beam intensity, especially for extracting the inner core of the beam, very small ripple for the power supplies of magnets is required. In this article, a different approach to extract beams from synchrotrons with higher beam stability is introduced.

A special multipole magnet structure was proposed earlier [5], which has the advantages of providing both symmetric and anti-symmetric high-order fields of same order and cheaper in cost due to the simplified structure. In addition, the similar structure can be applied to any order of multipole magnets. When studying the potential applications in beam dynamics with the special magnets, we found that it is interesting that anti-symmetric sextupole magnets have the potential to be used for resonant slow extraction from synchrotrons. It is found that an anti-symmetric second-order field can shrink effectively the stable region area in the phase plane when the working point is near a half-integer resonance [6]. This property can be used to extract particles when one either tunes the resonance strength or blows up the emittance just like in the case of the usual third-order resonant slow extraction method by a standard sextupole. Our studies show that an anti-symmetric sextupole could produce a stronger perturbation to the beam dynamics in a larger tune range than a standard sextupole does, which could benefit the slow extraction. The beam intensity extracted by this method is more stable and less affected by working point jitters than that by the usual third-order resonant method.

Furthermore, the method to excite resonances by using anti-symmetric sextupoles might have other applications. For example, it can be considered to remove beam halo in very large proton-proton colliders such as FCC-hh [7] and SPPC [8]. One can shrink the acceptance without moving the work point much close to the half-integer resonance, and the stable region is similar in shape to the initial beam distribution in phase space. This means that it is easy to restore to the nominal operation by taking away the anti-symmetric sextupole field. These merits might be useful for one to swing between the collimation mode and the normal collision mode.

**2. Resonant slow extraction method by anti-symmetric sextupole**

*2.1. Anti-symmetric sextupole magnets*

In the current-free region of a magnet gap, the field can be derived from differential of the scalar potential. For a standard sextupole, the magnetic field can be expressed as [9],

$$B_x = \left(\frac{d^2 B_y}{dx^2}\right)_0 xy$$

$$B_y = \frac{1}{2}\left(\frac{d^2 B_y}{dx^2}\right)_0 (x^2 - y^2). \tag{1}$$

Under normal circumstances, the distribution of magnetic field produced by standard multipole magnets is fixed, e.g. the magnetic field $B_y$ of even order is a symmetric distribution about the horizontal plane and the one of odd order is an anti-symmetric distribution. However, if a pair of magnetic field shielding plates is placed at the center of two pairs of poles to decouple the two halves of the magnets, one can obtain any order either symmetric or anti-symmetric multipole field distribution through adjusting the shape of the poles and the dimensions and positions of the shielding plates. The relative field errors of about 1% in a good field region of rectangular shape can be obtained, which is considered acceptable for most applications. Fig. 1 shows the layout and magnetic field flux of an anti-symmetric sextupole.

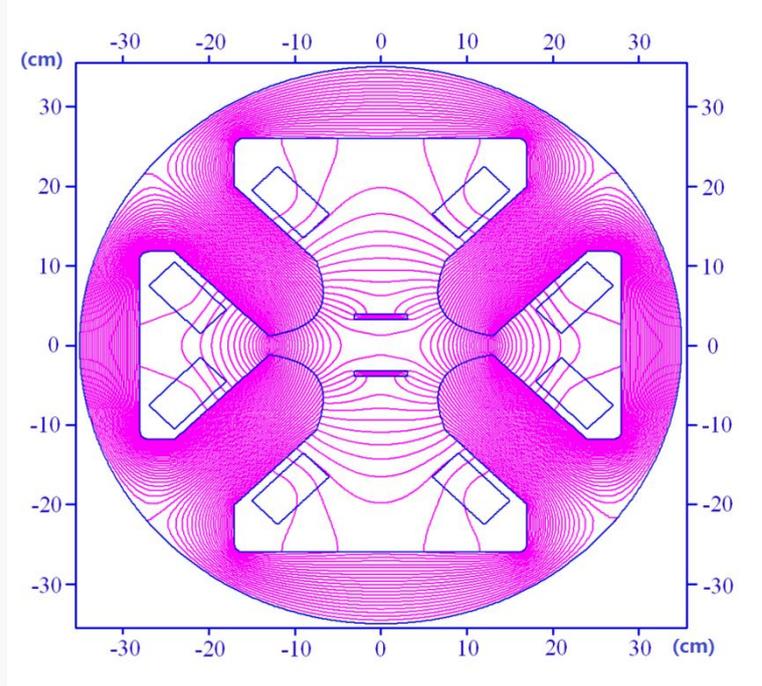

Fig. 1. Layout and magnetic field flux of an anti-symmetric sextupole

Similar to a standard sextupole, the magnetic field produced by an anti-symmetric sextupole magnet can be expressed as,

$$B_x = \left(\frac{d^2 B_y}{dx^2}\right)_0 |x|y$$

$$B_y = \frac{1}{2}\left(\frac{d^2 B_y}{dx^2}\right)_0 (x^2 - y^2)\frac{|x|}{x}. \tag{2}$$

As mentioned in Ref [9], this kind of field distribution is not realistic but a good approximation. In a real case, the field distribution is connected by a much more complicated form to connect the two sides of the Y-axis. As the difference between the realistic distribution and Eq. (2) or the error that is discussed in Section 3.3 is relatively small, such treatment is acceptable for most applications. It is the same situation when the Hamiltonian is introduced in Section 2.2.

In most cases where the magnet length is not very long, the effect of a multipole field on particle trajectories can be described in a simple way just by treating the magnet as a thin lens. For a positively charged particle in an anticlockwise ring, the effect of a thin-lens anti-symmetric sextupole in normalized co-ordinates appears as,

$$\Delta X = \Delta Y = 0$$

$$\Delta X' = \beta_x^{1/2} \frac{B_y l_s}{|B\rho|} = S\left(X^2 - \frac{\beta_y}{\beta_x} Y^2\right) \frac{|X|}{X}$$

$$\Delta Y' = -2S \frac{\beta_y}{\beta_x} |X| Y, \tag{3}$$

where $S$ is the normalized sextupole strength,

$$S = \frac{1}{2} \beta_x^{3/2} \frac{l_s}{|B\rho|} \left(\frac{d^2 B_y}{dx^2}\right)_0. \tag{4}$$

According to Eq. (3), one can see that just as other multipole magnets an anti-symmetric sextupole couples the horizontal and vertical motions unless $Y = 0$ and the strength of the coupling is proportional to the ratio of the vertical and horizontal betatron amplitude functions ($\beta_y/\beta_x$). For extraction in the horizontal plane, provided the vertical tune does not satisfy the resonance condition, one places the magnet at a location where $Y$ is much smaller than $X$, so we can neglect the influence of the vertical motion. Eq. (3) can be rewritten with a much simplified form,

$$\Delta X = \Delta Y = \Delta Y' = 0$$

$$\Delta X' = S|X|X. \tag{5}$$

According to Eq. (5), one can see that when a positively charged particle passes through a thin-lens anti-symmetric sextupole, only the horizontal angle has been changed if ignoring the influence of the vertical motion and the change value is relevant to the sextupole strength and the horizontal position of the particle.

### 2.2. Kobayashi Hamiltonian

As we know, standard sextupole may cause third-order resonance when the betatron tune is close to a third-integer. Now we will derive the Kobayashi Hamiltonian [10] close to resonance lines to see whether an anti-symmetric sextupole could also cause resonance.

In synchrotrons, from the general transfer matrix in the normalized co-ordinates,

$$M_n = \begin{pmatrix} \cos 2\pi(nQ_x) & \sin 2\pi(nQ_x) \\ -\sin 2\pi(nQ_x) & \cos 2\pi(nQ_x) \end{pmatrix}. \tag{6}$$

First with the horizontal betatron tune close to a half-integer, i.e. $Q_x = m+1/2+\delta Q$, one can obtain the transfer matrix of the particle after one and two turns, where $m$ is integer and $|\delta Q| \ll 1/2$,

$$M_1 = \begin{pmatrix} -1 & -\varepsilon \\ \varepsilon & -1 \end{pmatrix}, \quad M_2 = \begin{pmatrix} 1 & 2\varepsilon \\ -2\varepsilon & 1 \end{pmatrix}, \tag{7}$$

where $\varepsilon$ replaces $2\pi\delta Q$ for brevity. From Eq. (7) it can be seen that a particle at the exact half-integer resonant tune will return to its initial position every two turns. Now we will calculate the change of the position and divergence of the particle after two turns with the anti-symmetric sextupole as a perturbation by the linear addition. The particle will pass through the anti-symmetric sextupole after each revolution and after two turns the coordinates of the particle could be written as,

$$\begin{pmatrix} X_2 \\ X_2' \end{pmatrix} = N_2 M_1 N_1 M_1 \begin{pmatrix} X_0 \\ X_0' \end{pmatrix}, \tag{8}$$

where the matrix $M_1$, $N_1$, and $N_2$ denote $\begin{pmatrix} -1 & -\varepsilon \\ \varepsilon & -1 \end{pmatrix}$, $\begin{pmatrix} 1 & 0 \\ S|X_1| & 1 \end{pmatrix}$, $\begin{pmatrix} 1 & 0 \\ S|X_2| & 1 \end{pmatrix}$, respectively.

Then we can obtain the expressions, known as the spiral step and spiral kick:

$$\Delta X_2 = X_2 - X_0 = 2\varepsilon X_0'$$
$$\Delta X_2' = X_2' - X_0' = -2\varepsilon X_0 + SX_0|X_0| + S(X_0 + \varepsilon X_0')|X_0 + \varepsilon X_0'|. \tag{9}$$

As $\varepsilon$ is a small quantity when the tune is close to a half-integer, and $X_0$ and $X'_0$ are also small, higher-order terms about $\varepsilon$, $X_0$ and $X'_0$ could be neglected. For example, one can also neglect the item $\varepsilon SXX'$ which may cause the stable region a slight incline. Then Eq. (9) could be written as,

$$\Delta X_2 = X_2 - X_0 = 2\varepsilon X_0'$$
$$\Delta X_2' = X_2' - X_0' = -2\varepsilon X_0 + 2SX_0|X_0|. \tag{10}$$

The time needed for two revolutions in the machine is very short compared to the beam spill time during extraction, so it can be safely used as the basic time unit. Thus the subscripts are no longer needed and Eq. (10) could be treated as a continuous function that is derived from a Hamiltonian,

$$\Delta X_2 => \left(\frac{\Delta X}{\Delta t}\right)_{\Delta t=1(2\ turn)} = \dot{X} = \frac{\partial H}{\partial X'} = 2\varepsilon X'$$

$$\Delta X_2' => \left(\frac{\Delta X'}{\Delta t}\right)_{\Delta t=1(2\ turn)} = \dot{X}' = -\frac{\partial H}{\partial X} = -2\varepsilon X + 2SX|X|. \tag{11}$$

Finally one can obtain the Kobayashi Hamiltonian for an anti-symmetric sextupole working at tune close to a half-integer resonant line. By dealing with the absolute value sign and integrating the partial differentials, they are expressed as

$$X \geq 0 \quad H = \varepsilon(X^2 + X'^2) - \frac{2}{3}SX^3$$

$$X < 0 \quad H = \varepsilon(X^2 + X'^2) + \frac{2}{3}SX^3, \tag{12}$$

Fig. 2 shows the phase-space map calculated from the Kobayashi Hamiltonian. It can be seen that if $\varepsilon/S > 0$ (in the figure, $S > 0$ is used), the particle trajectories in the phase space with the anti-symmetric sextupole will have a stable region and the motions of the particles out of it are unstable. As the tune is moved closer to the half-integer resonance or strength of the anti-symmetric sextupole is increased, the area of stable region shrinks gradually just as in the case of the standard sextupole at the third-order resonance. However, one can see that if $\varepsilon/S < 0$, there is no unstable region even if the tune is close enough to the half-integer or the strength of the anti-symmetric sextupole is large enough. This is quite different from the usual third-order resonance that has symmetric maps on both sides of the resonance. The reason can be explained as follows: assuming that $\varepsilon < 0$ and $S > 0$, from Eq. (10) one can see that if a particle is in the first quadrant in the phase-space plane, $\Delta X_2$ is always less than zero and $\Delta X'_2$ is always greater than zero. That is to say after some turns this particle must be in the second quadrant. Similarly, it will be in the third quadrant and fourth quadrant and then in the first quadrant again. This means that the particle will cycle in the phase plane and cannot be extracted from the ring. Therefore, with an anti-symmetric sextupole the resonance happens only in one side of a half-integer. Besides, from the phase-space map one can see that with an anti-symmetric sextupole the stable region has two separatrices as compared to three separatrices in the case of a standard sextupole at the third-order resonance. Even more, here the stable region is less distorted and close to the initial distribution.

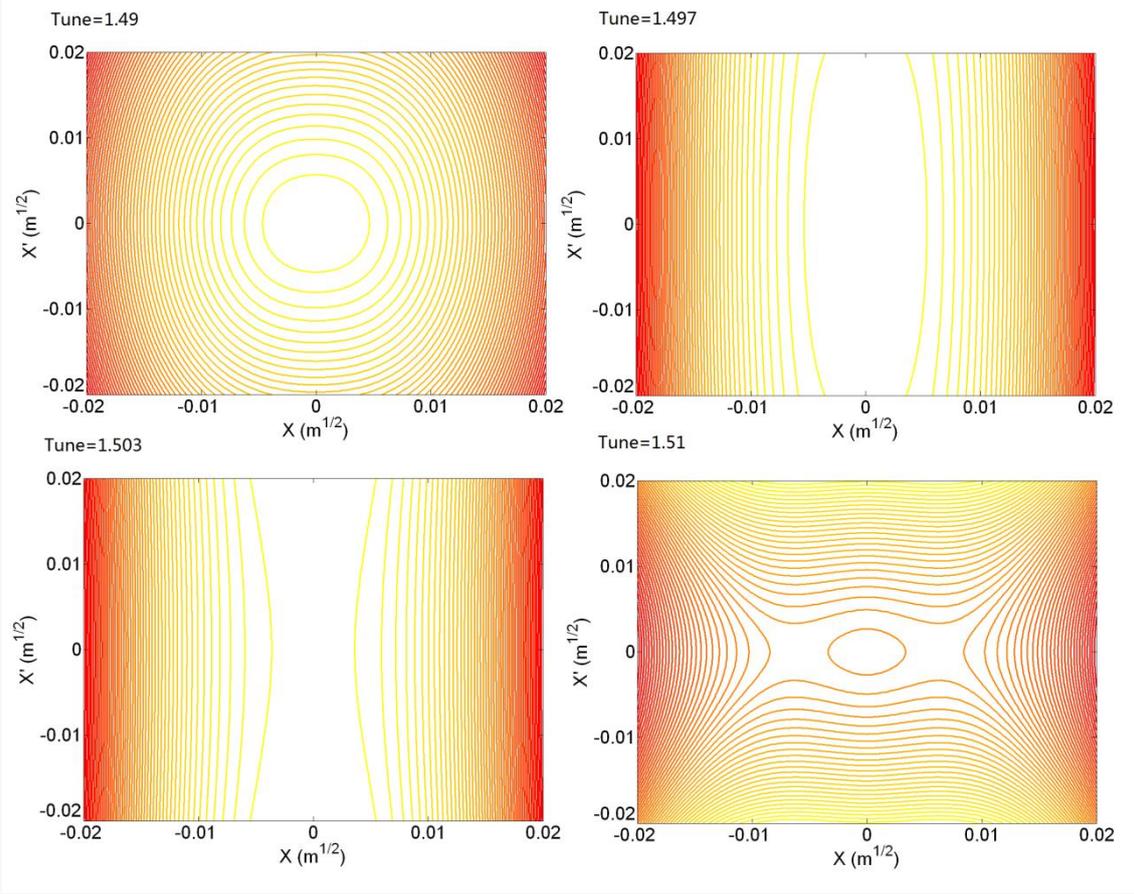

Fig. 2. Phase-space maps (or Poincaré maps) for the horizontal betatron tune crossing a half-integer from left

With the same derivation process, we could obtain the Kobayashi Hamiltonian for the tune close to one third of an integer with an anti-symmetric sextupole,

$-\frac{1}{2}X - \frac{\sqrt{3}}{2}X' > 0$, and $-\frac{1}{2}X + \frac{\sqrt{3}}{2}X' > 0$,  $H = \frac{3\varepsilon}{2}(X^2 + X'^2) + \frac{1}{3}SX^3 + S(\frac{3}{4}XX'^2 + \frac{1}{12}X^3)$

$-\frac{1}{2}X - \frac{\sqrt{3}}{2}X' < 0$, and $-\frac{1}{2}X + \frac{\sqrt{3}}{2}X' < 0$,  $H = \frac{3\varepsilon}{2}(X^2 + X'^2) - \frac{1}{3}SX^3 - S(\frac{3}{4}XX'^2 + \frac{1}{12}X^3)$

$-\frac{1}{2}X - \frac{\sqrt{3}}{2}X' > 0, -\frac{1}{2}X + \frac{\sqrt{3}}{2}X' < 0$, and X>0,  $H = \frac{3\varepsilon}{2}(X^2 + X'^2) - \frac{1}{3}SX^3 + \frac{\sqrt{3}}{4}S(X^2X' + X'^3)$

$-\frac{1}{2}X - \frac{\sqrt{3}}{2}X' > 0, -\frac{1}{2}X + \frac{\sqrt{3}}{2}X' < 0$, and X<0,  $H = \frac{3\varepsilon}{2}(X^2 + X'^2) + \frac{1}{3}SX^3 + \frac{\sqrt{3}}{4}S(X^2X' + X'^3)$

$-\frac{1}{2}X - \frac{\sqrt{3}}{2}X' < 0, -\frac{1}{2}X + \frac{\sqrt{3}}{2}X' > 0$, and X>0,  $H = \frac{3\varepsilon}{2}(X^2 + X'^2) - \frac{1}{3}SX^3 - \frac{\sqrt{3}}{4}S(X^2X' + X'^3)$

$-\frac{1}{2}X - \frac{\sqrt{3}}{2}X' < 0, -\frac{1}{2}X + \frac{\sqrt{3}}{2}X' > 0$, and X<0,  $H = \frac{3\varepsilon}{2}(X^2 + X'^2) + \frac{1}{3}SX^3 - \frac{\sqrt{3}}{4}S(X^2X' + X'^3)$  (13)

and also the Kobayashi Hamiltonian for the tune close to one quarter of an integer,

$X \geq 0$, and $X' \geq 0$, $\qquad H = 2\varepsilon(X^2 + X'^2) - \frac{2}{3}S(X^3 + X'^3)$

$X \geq 0$, and $X' < 0$, $\qquad H = 2\varepsilon(X^2 + X'^2) - \frac{2}{3}S(X^3 - X'^3)$

$X < 0$, and $X' \geq 0$, $\qquad H = 2\varepsilon(X^2 + X'^2) + \frac{2}{3}S(X^3 - X'^3)$

$X < 0$, and $X' < 0$, $\qquad H = 2\varepsilon(X^2 + X'^2) + \frac{2}{3}S(X^3 + X'^3).$ (14)

Figs. 3 and 4 show the phase-space maps calculated from the Kobayashi Hamiltonian expression for the tune close to one third of an integer and one quarter of an integer, respectively. It can be seen that the anti-symmetric sextupole also has property to drive single-side higher-order resonances, similar to the case of the half-integer resonance. However, those higher-order resonances cannot be used to extract particles as there are super stable regions outside the unstable areas.

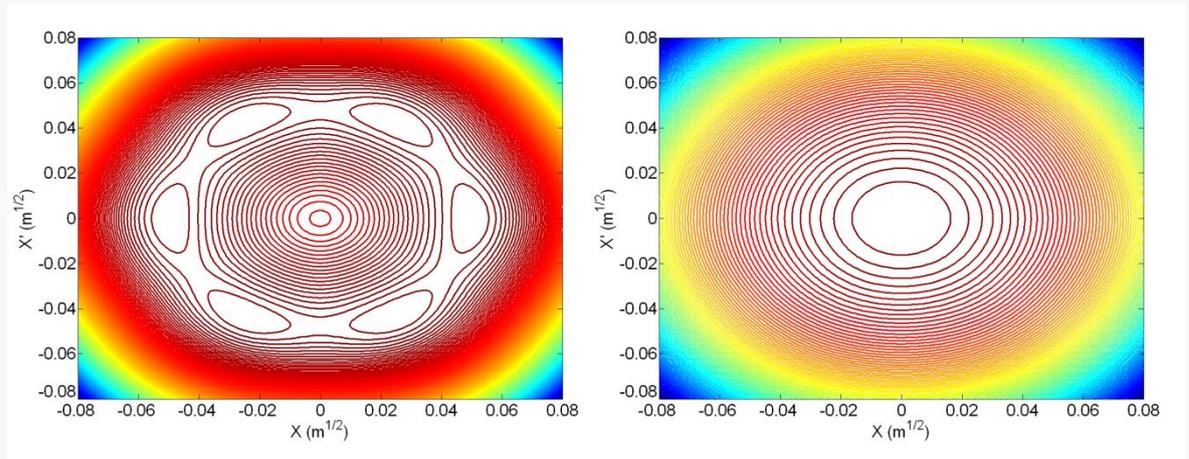

Fig. 3. Phase-space map calculated from the Kobayashi Hamiltonian for the tune close to one third of an integer with an anti-symmetric sextupole (Left: $\varepsilon/S > 0$; Right: $\varepsilon/S < 0$)

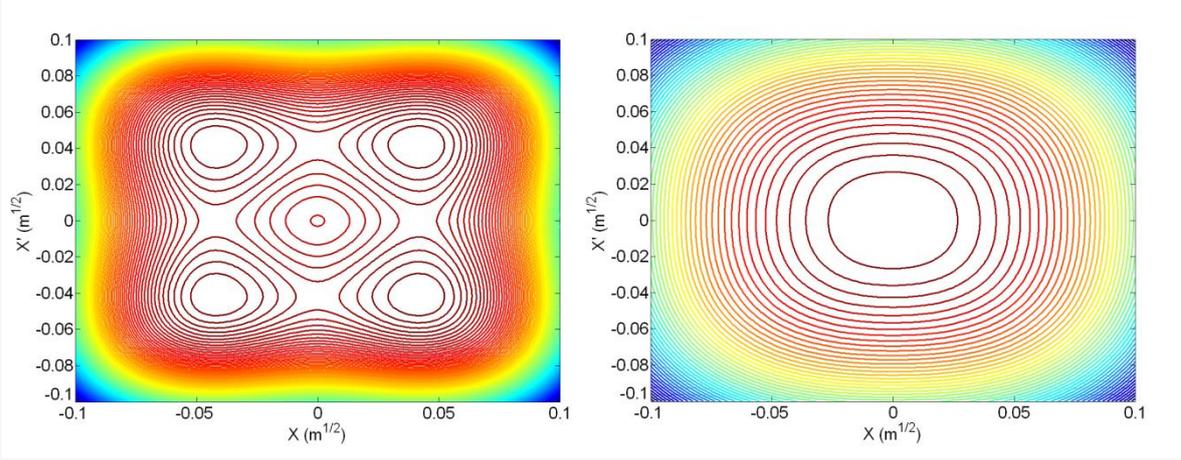

Fig. 4. Phase-space map calculated from the Kobayashi Hamiltonian for the tune close to one quarter of an integer with an anti-symmetric sextupole (Left: $\varepsilon/S > 0$; Right: $\varepsilon/S < 0$)

*2.3. Stable region size at half-integer resonance with an anti-symmetric sextupole*

The area of the stable region in the phase space at third-integer resonance with a standard sextupole (SX) can be expressed as follows [9],

$$A_{SX} = 48\sqrt{3}\ \pi^2 \left(\frac{\delta Q}{S}\right)^2. \tag{15}$$

Similarly, through Kobayashi Hamiltonian the area of the stable region at half-integer resonance with an anti-symmetric sextupole (ASX) can also be expressed by an empirical formula,

$$A_{ASX} = 5.905\pi^2 \left(\frac{\delta Q}{S}\right)^2. \tag{16}$$

The empirical formula was obtained from the phase-space maps calculated from the Kobayashi Hamiltonian. It is found that with different phase-space maps, the calculated coefficients are highly consistent, e.g. the error is less than $10^{-5}$. From Eqs. (15) and (16), in both cases, the stable region is proportional to $(\delta Q/S)^2$. Figs. 5 and 6 show the stable region areas for the two cases with respect to the sextupole strength and the tune distance to its resonance, respectively. As indicated by the coefficients in Eqs. (15) and (16), the area of stable region caused by anti-symmetric sextupole is about 1/14 of the one by standard sextupole. In other words, to have the same area of stable region, together with the same sextupole strength, the tune distance to resonance in the case of anti-symmetric sextupole is $\sqrt{14}$ times the one in the case of standard sextupole, and this represents the advantage of less sensitivity to the tune variation due to the ripple of power supplies of magnets. To be specific, with the same sextupole strength and the same tune distance to the resonance, the change of the stable region area due to the tune variation or *dA/dδQ* in the case of anti-symmetric sextupole is 1/14 of the one in the case of standard sextupole.

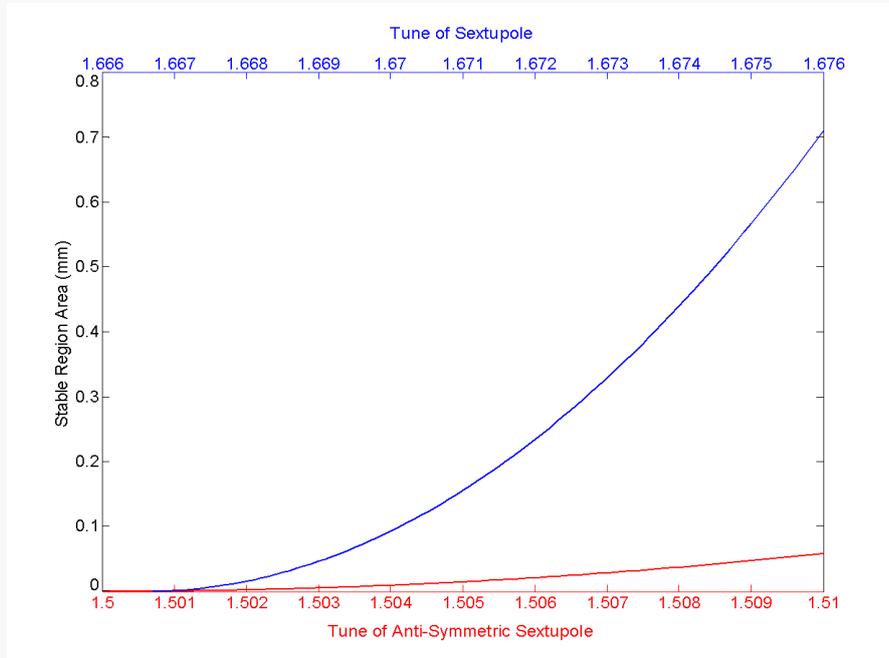

Fig. 5. Stable region areas versus the same sextupole strength ($S = 10$ m$^{-1/2}$) in the cases of SX and ASX

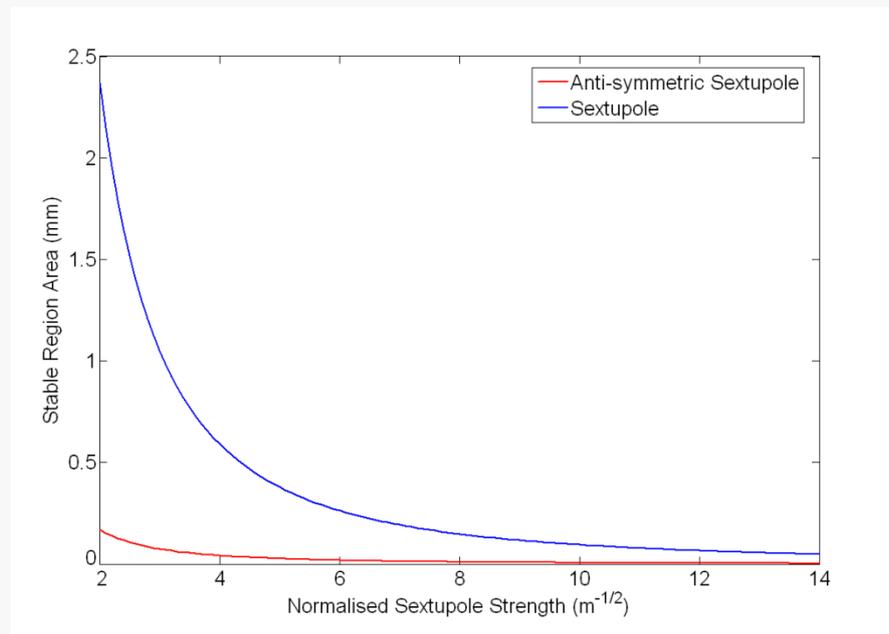

Fig. 6. Stable region areas versus the tune distance to the related resonance ($\delta Q = 0.0034$) in the cases of SX and ASX

*2.4. Resonant slow extraction driven by anti-symmetric sextupoles*

Same as in the case of the resonant slow extraction by a standard sextupole, the formation and separatrices of the stable region by an anti-symmetric sextupole at a tune close to a half-integer can also be used to extract beams from synchrotrons. After the particles escape from the stable region, they will follow one of the separatrices to move outward. Before they get lost in the vacuum chamber, a well-positioned electrostatic septum (ES) will guide them into the extraction channel.

If one takes into account the spiral step and the positions of the ES and the septum magnet, the whole slow extraction becomes more complicated. On one hand, to have a lower requirement on the ES strength, one needs to produce a phase advance close to 90° between the ES and the septum magnet. On the other hand, one wishes to have a smaller angle of the extraction path to the X-axis (hereafter, angle always means with respect to the X-axis) to make the spiral step more effective at the ES. However, in the case of the third-order resonant extraction, the angle for the extraction path at the septum magnet is limited by the other extraction arms of the beam, and should be larger than -60° (assuming extraction from the outside of the ring); for the same reason the angle for the extraction path at the ES is also limited to be less than 60°. This means that the angle at the ES is 45±15°, and the one at the septum magnet is 45±15°. In the case of the new method, one has much more freedom to place the ES and the septum magnet, as there are only two extraction arms. By the way, the half-integer resonant extraction with octupole has also such freedom. One can place the ES at a smaller angle, e.g. 10°, and the septum magnet can be placed 70-90° downstream depending on the lattice layout. For the step size at the ES which is related to the beam loss at the ES wires, in general, with the same strength for the driving magnet, the spiral step is a little smaller for the new method as it has only two driven turns compared to three turns for the traditional method. However, if you take into account that one can use smaller angle at the ES, the difference is small. As the extraction arm has an angle of about 60° at the anti-symmetric sextupole, the ES is better placed with a phase advance of about 50° downstream the magnet.

One can estimate the outward step after two turns with anti-symmetric sextupole as follows: according to Eq. (10), taking $v_x = 1.512$, $S = 10$ m$^{-1/2}$, the $X$ position at the ES is 0.03 m$^{1/2}$, the horizontal betatron function at the ES (same location as the driving magnet) is 6.46 m, and the normalized divergent angle ($X'$) is 0.02 m$^{1/2}$, one can obtain $\Delta X_2 \approx 7.6$ mm. It is much larger than the width of an ES septum which is usually about 0.1 mm and the extraction efficiency will be sufficiently high. As described above, the step size can be enhanced by placing the ES to a downstream location.

For the usual third-order resonance extraction method, when the beam core is to be extracted, the stable region should be shrunk by either moving the tune closer to the resonance or increasing the driving force (sextupole strength). In recent decades, the method using fixed resonance but blowing up the horizontal beam emittance by a so-called RF knock-out (or RFKO) system, or the RFKO method, has been developed [11-16]. This method has the advantages: the tune is fixed at a relatively larger distance from the resonance, which means that the extraction process is less sensitive to the tune variation due to the ripple of power supplies; the resonance can be paused in the course of extraction; more homogeneous beam intensity can be obtained. We can employ the same RFKO method in the resonant slow extraction by anti-symmetric sextupoles. In this case, the fixed tune can be placed at a relatively larger distance from a half-integer, and the inner core of the beam will be extracted gradually by applying the RFKO continuously. As the emittance is less deformed by anti-symmetric sextupoles, it is relatively easy to stop or suspend the extraction process by reducing the driving field strength.

For practical applications, one should also consider the chromaticity effect caused by momentum spread, which increases the beam loss at the ES slightly. As mentioned in Ref [9], one usually applies the Hardt condition to suppress the effect which causes the extraction path depending on the particle's momentum, if one wishes to minimize the loss. This can be assisted by chromaticity correction sextupoles. The anti-symmetric sextupole has also an influence to the chromaticity that is different from the standard sextupole and still under study.

*2.5. More rigorous mathematical treatment*

For the expression of Eq. (2), the field is non-continuous and non-differentiable at $x = 0$ if $y \neq 0$. Although it is claimed that the expression is only a good approximation to a real magnetic field distribution, it is better to show that this non-continuity will not produce serious problem to the physical image. To solve this problem partially, for a given example: $\left(\frac{d^2 B_y}{dx^2}\right)_0 = 24.32$ G/cm², the good field range $-10\,\text{cm} < x < 10\,\text{cm}$ and $-1\,\text{cm} < y < 1\,\text{cm}$, the following expression is found to be a good approximation to represent the middle region between the two halves of the magnet:

$$B_y(x, y) = \begin{cases} 12.16 x|x| - 6.64 xy^2, & |x| \leq 1.83\,\text{cm} \\ 12.16(x^2 - y^2)x/|x|, & |x| > 1.83\,\text{cm} \end{cases}, \tag{17}$$

$$B_x(x,y) = \begin{cases} 24.32|x|y - 2.21y^3, & |x| \le 1.83\,\text{cm} \\ 24.32|x|y, & |x| > 1.83\,\text{cm} \end{cases}$$, (18)

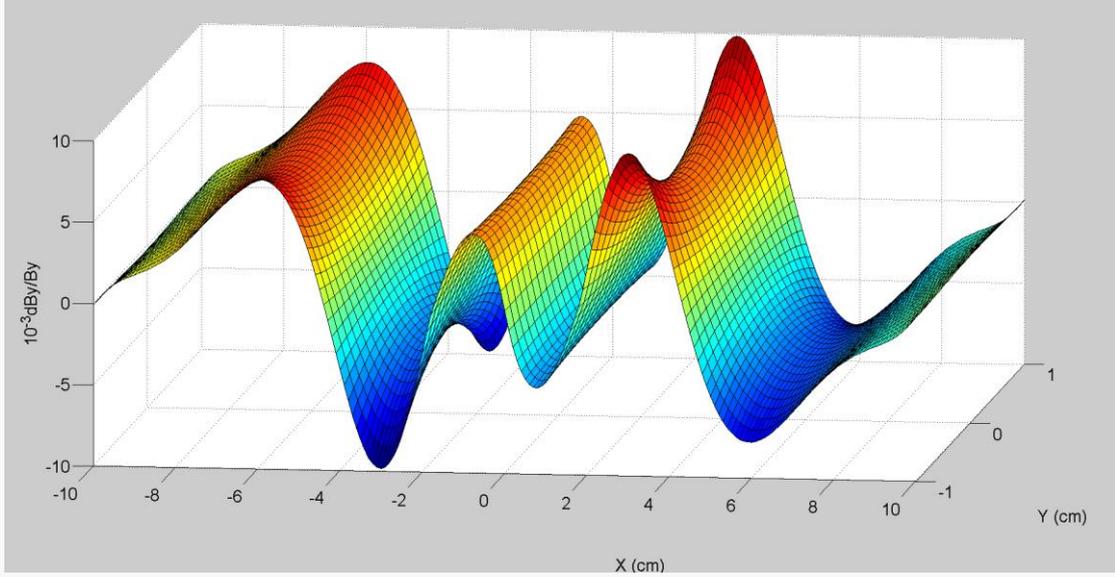

Fig. 7. Difference errors on $B_y$ between the composed field and the calculated field

Fig. 7 shows the difference errors on $B_y$ between the composed field and the calculated field, and this can be compared to Figure 18 in Section 3.3. One can find that with this composed field we solve the problem of $B_y$ discontinuity at $x = 0$. However, at the intersections $x = -1.83$ cm and $x = 1.83$ cm, the horizontal field component is made non-continuous. As the horizontal field component has little influence to the particle's motion at this special location of very small $\beta_y$, this drawback is acceptable.

The 2-D tracking results show that there is no evident difference between the ideal anti-symmetric sextupole and the composed field distribution. Together with the numerical tracking results using calculated 2-D field distribution to be shown in Section 3, this approach confirms that the approximate expression of anti-symmetric sextupole for a realistic magnetic field is reasonable. For simplicity, for the rest part of the article, standard anti-symmetric sextupole is used for the ideal field.

### 3. Multi-particle simulations

*3.1. Simulation/calculation codes and initial conditions*

Multi-particle simulations have been carried out to confirm the method in realistic cases. Because it is not easy to add anti-symmetric sextupole fields into the widely-used multi-particle simulation codes, a self-made code based on MATLAB has been developed. The code was benchmarked by testing the resonant slow extraction with standard sextupole fields and comparing with the WinAgile code. Another code also based on MATLAB and Kobayashi Hamiltonian method has been developed to calculate the phase space maps.

As an example, a synchrotron lattice for proton therapy (APTF) [17] was used for the simulation study, and the initial beam distribution is a 2-D Gaussian distribution in both horizontal and vertical phase plane. The initial rms emittance is 10 $\pi$mm mrad in both planes with $\beta_x$=1.30 m, $\beta_y$ = 8.47 m, $D_x$ = 1.58 m and the initial particle number is 20000. The anti-symmetric sextupole is placed at Sextupole 3 where one has $\beta_x$ = 6.46 m, $\beta_y$ = 1.51 m, $D_x$ = 2.7 m. The momentum spread of $\pm 10^{-3}$ is chosen and included in the simulations. For a natural chromaticity of $\xi_x$ = -1.0, the tune spread due to the momentum spread is about 0.001, almost negligible in this case where the tune distance is large. Fig. 8 shows the layout, and the betatron and dispersive functions of the synchrotron lattice. Fig. 9 shows the initial beam distributions. The simulations were carried out first with 1-D field distribution, then 1-D and 2-D inherent field errors.

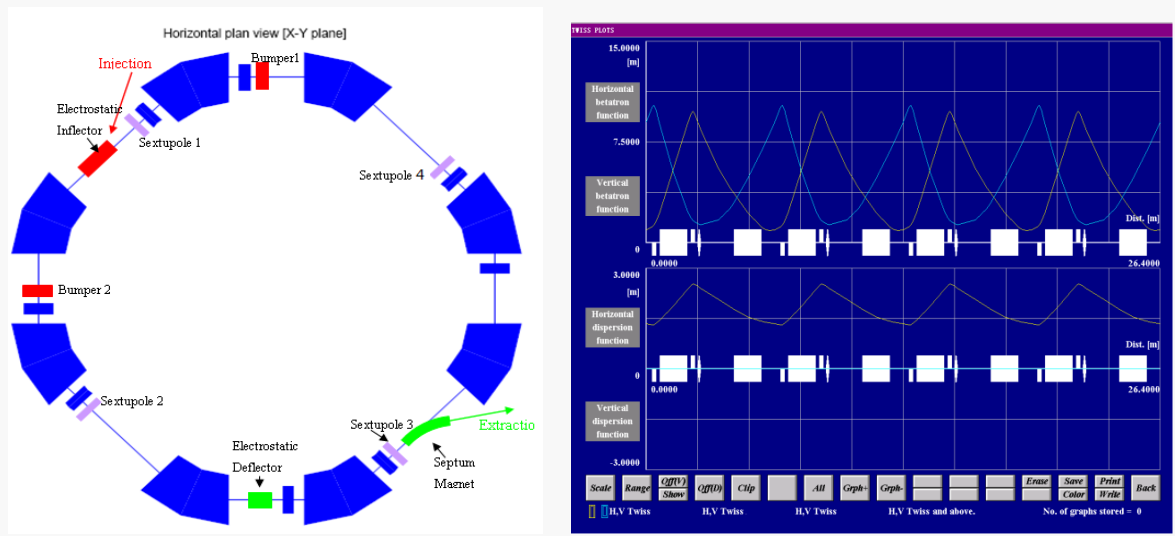

Fig. 8. Layout and betatron function of the synchrotron which is used for the simulations using an anti-symmetric sextupole (Left: layout of the synchrotron; Right: betatron and dispersive functions)

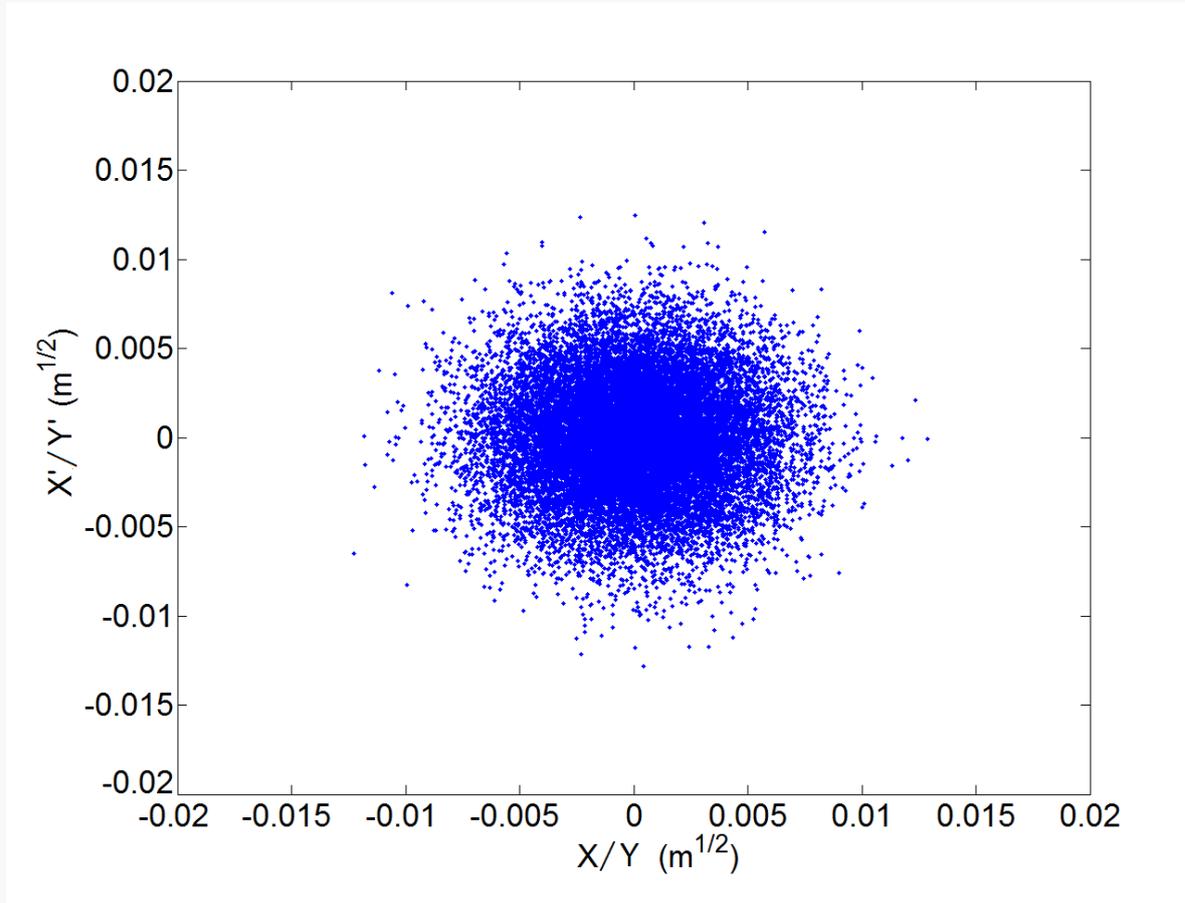

Fig. 9. Initial beam distribution (2-D Gaussian) in the normalized phase space for simulations

*3.2. Simulation results*

The self-made code was used to simulate the resonant slow extraction by using anti-symmetric sextupole fields. Fig. 10 shows the phase-space distributions which reflect the stable regions with anti-symmetric sextupole fields at different tunes close to a half-integer. Fig. 11 shows the phase-space maps at different tunes calculated by using the Kobayashi Hamiltonian. From the two figures, one can see that the simulation results are consistent to the Hamiltonian results, and the detailed difference is given later in this section. To compare the resonance properties driven by an anti-symmetric sextupole and a standard sextupole, Fig. 12 shows the phase space distributions before, in the course of and after applying the driving fields with the same strength and same tune distance to resonance. One can see that the stable region caused by an anti-symmetric sextupole is smaller and less distorted as compared to that caused by a standard sextupole, which is consistent to the results from the Kobayashi Hamiltonian. One can also compare the extraction process by means of extracted particles or more precisely the particles escaped from the stable region, as shown in Fig. 13, where the same field strength and same tune distance to resonance are applied. With a smaller stable region in the case of anti-symmetric sextupole, one can extract more particles. Actually, in order to avoid the emittance dilution, for each simulation the sextupole strength should be increased slowly along with the resonant slow extraction process. To confirm the empirical formula Eq. (16), simulations were carried out and the comparison with the results by using the formula is illustrated in Fig. 14. One can see that the relative differences between the simulation and empirical formula results are less than 10% with different resonance strengths, which means that the empirical formula obtained from Kobayashi Hamiltonian method gives a good approximation to the description of the resonance. In fact, the empirical formula results are always smaller than the simulation results due to the approximation in the derivation process as in Eq. (10).

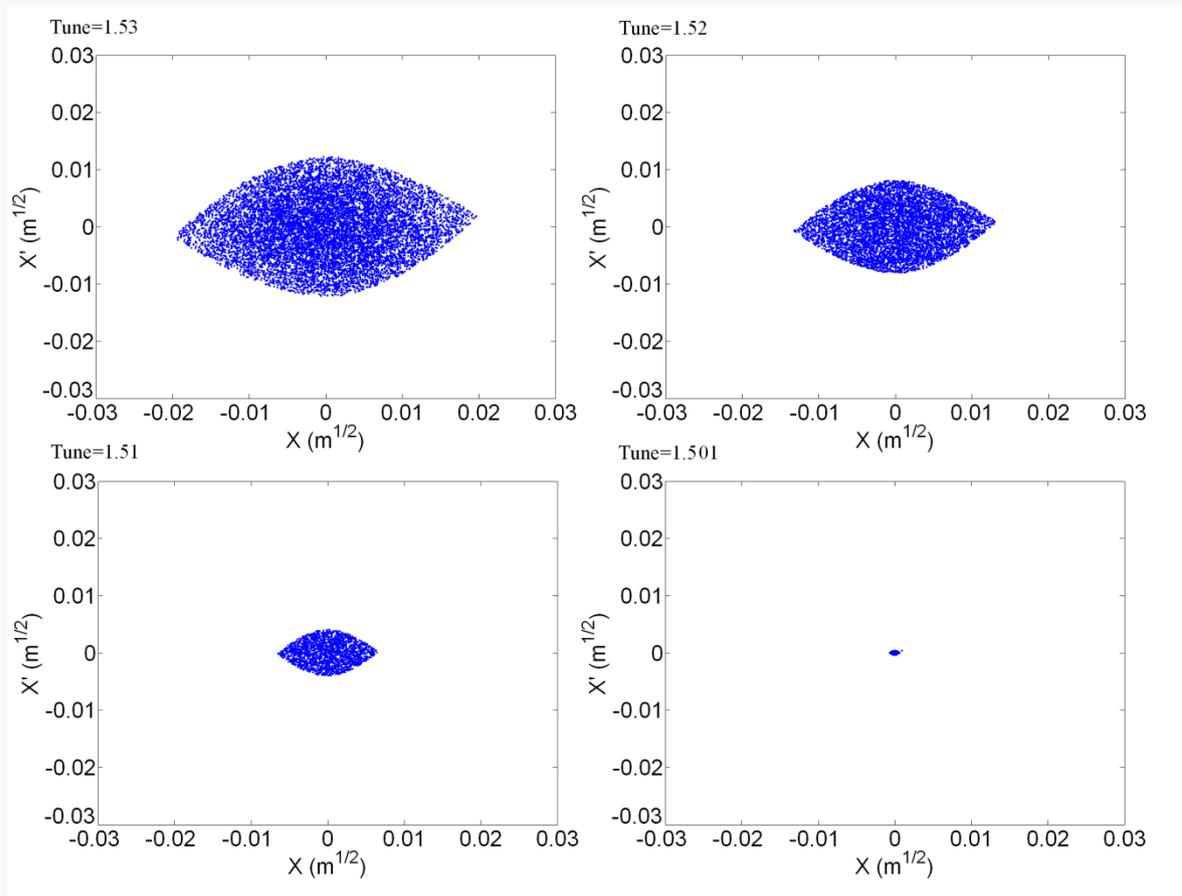

Fig. 10. Phase-space distributions with anti-symmetric sextupole fields and different tunes close to a half-integer

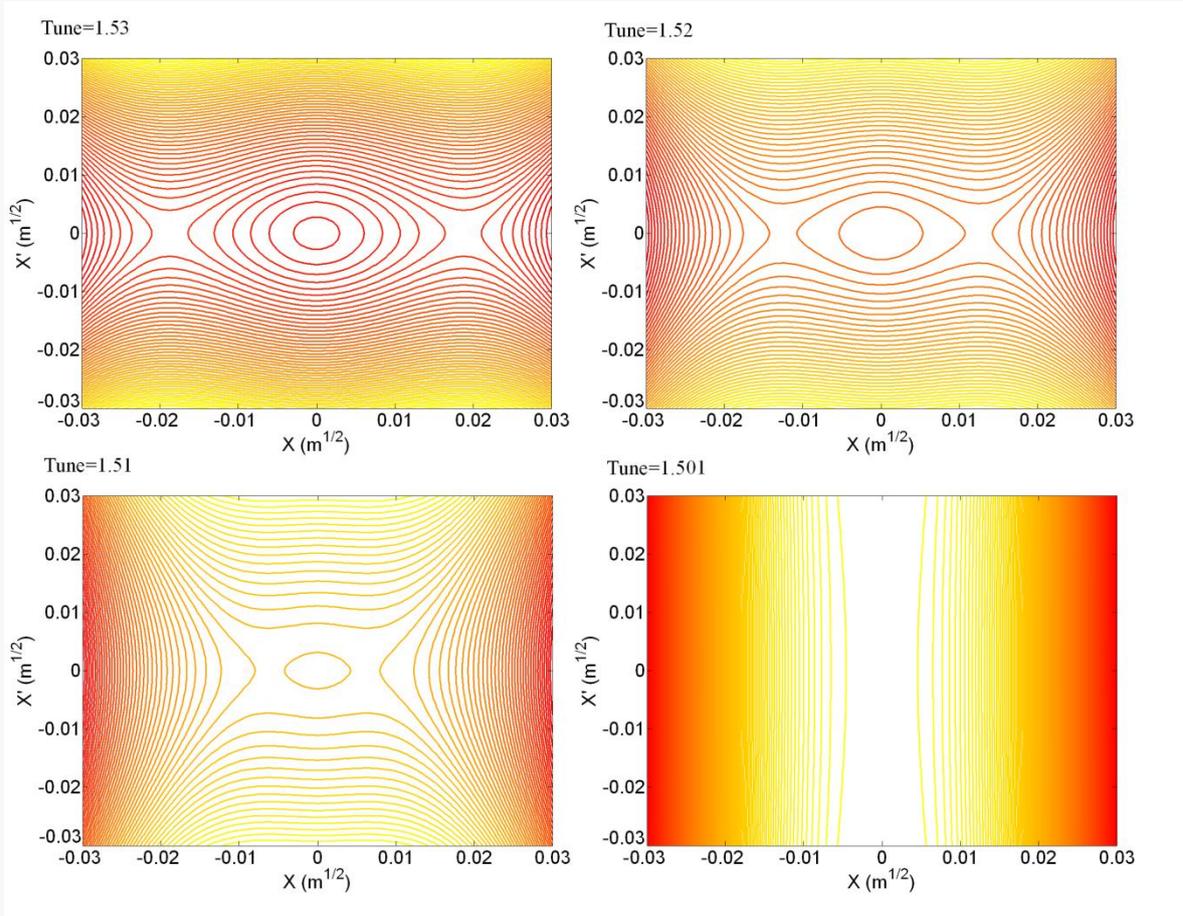

Fig. 11. Phase-space maps with different tunes close to a half-integer calculated from the Kobayashi Hamiltonian

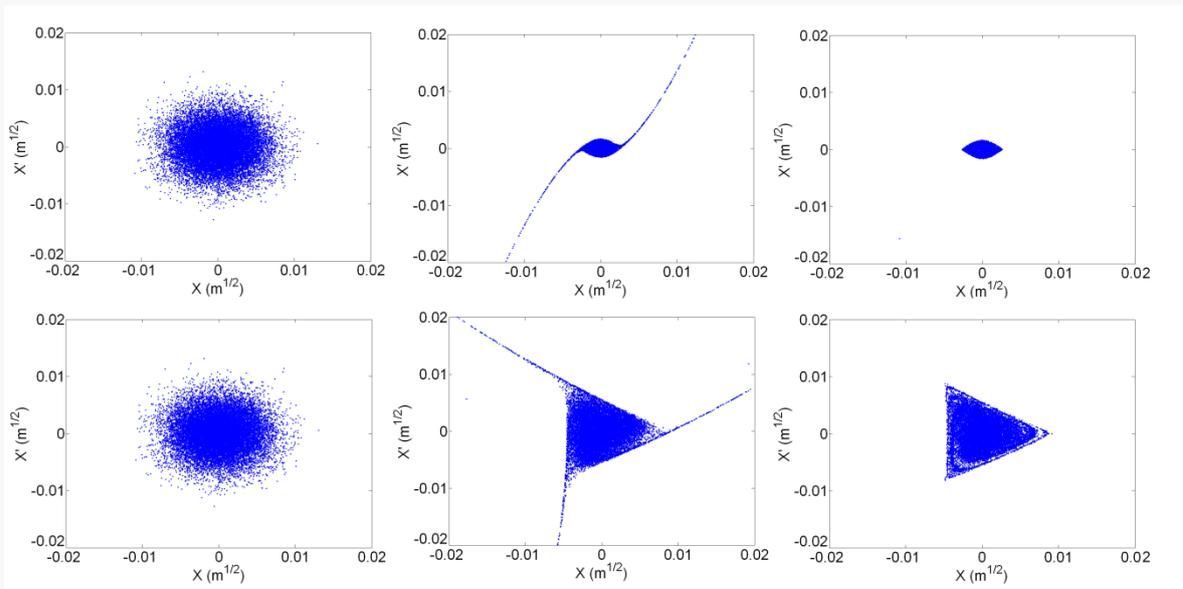

Fig. 12. Phase space distributions before, in the course of and after applying the driving fields (Upper: with an anti-symmetric sextupole; Lower: with a standard sextupole)

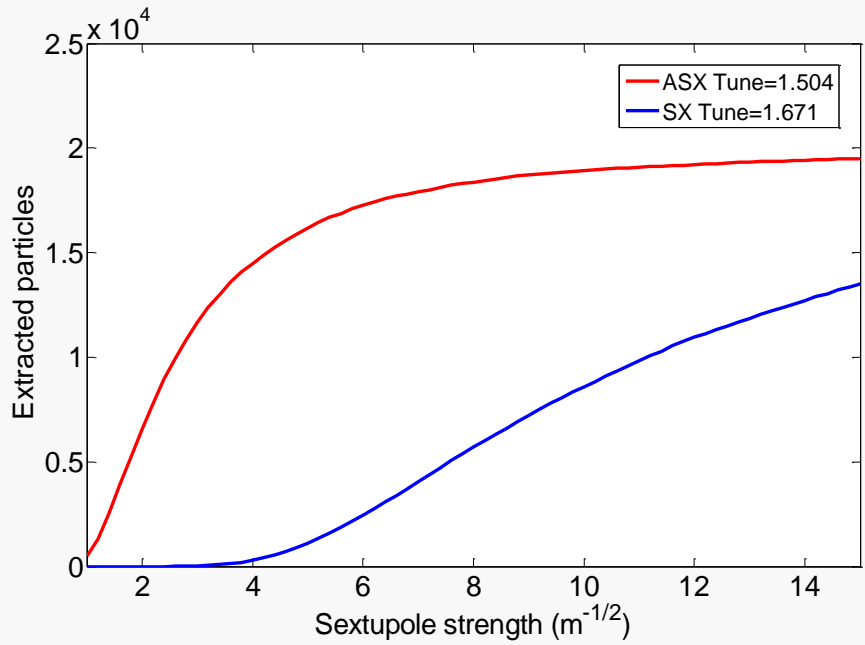

Fig. 13. The number of extracted particles by using anti-symmetric sextupole and standard sextupole with the same tune distance to resonance

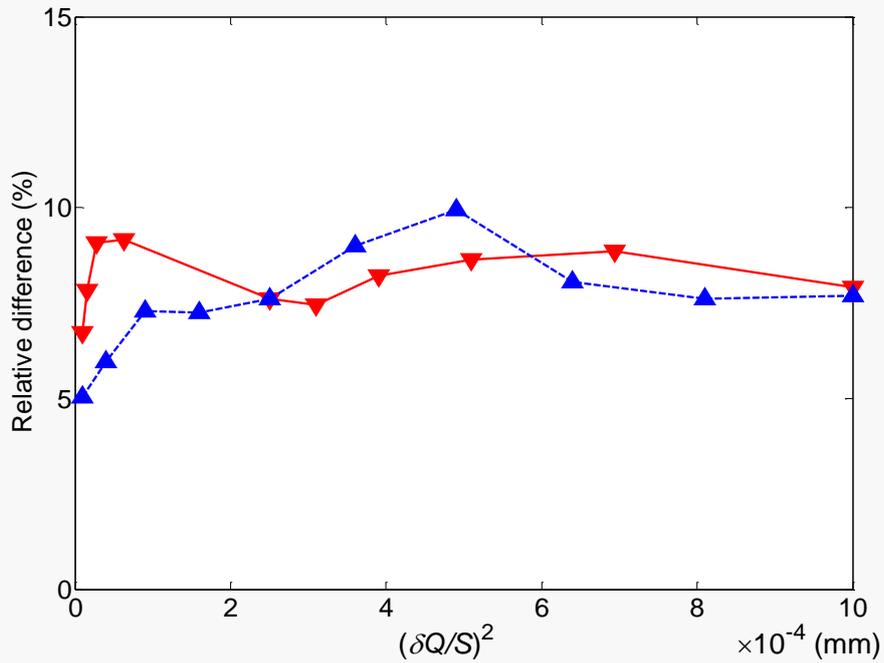

Fig. 14. Relative difference in the stable region areas between the simulation and empirical formula results with respect to different resonance strengths (Blue dotted line with regular triangles: with different tune distances from 0.001 to 0.01 and the field strength is 10 m$^{-1/2}$; red solid line with inverted triangles: with different field strengths from 50 m$^{-1/2}$ to 5 m$^{-1/2}$ and the tune distance is 0.005)

*3.3. Effect due to the inherent field error*

The results shown above are carried out with one-dimensional ideal sextupole fields. However, all magnets have field imperfection which is caused by different factors. The field errors will influence the beam dynamic behaviors. In addition, when including two-dimensional field distribution of a sextupole either standard or anti-symmetric, the beam behavior should be slightly different. As for an anti-symmetric sextupole, such as analyzed in Ref. [5], there is also additional inherent field error due to the special magnet structure, which are about 1% level depending on the magnet design. The so-called 1-D inherent field error means the difference error between the calculated field and the ideal one on the X-axis, and the 2-D inherent field error means the error in the X-Y plane. Here the influence to the resonant slow extraction by the field error will be discussed.

*3.3.1. 1-D inherent field error*

The 1-D inherent field error of anti-symmetric sextupole and its influence to the resonant slow extraction are discussed at first. As an example, Fig. 15 shows the 1-D inherent field error distribution in the horizontal median plane, which is from the two-dimensional magnetic field calculation of an anti-symmetric sextupole. Certainly the error distribution follows the anti-symmetric structure. The Kobayashi Hamiltonian including the field error is then derived.

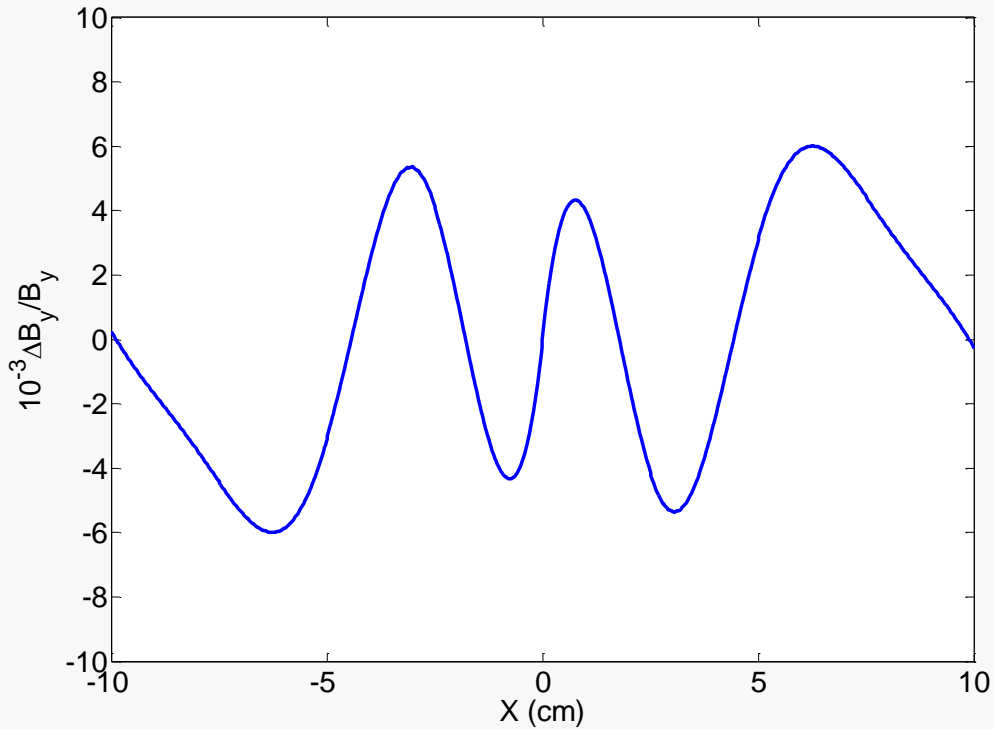

Fig. 15. Distribution of 1-D field error along the horizontal median plane for an anti-symmetric sextupole

By adding the field error $\Delta B_y$ to Eq. (3) one can obtain,

$$\Delta X = \Delta Y = 0$$

$$\Delta X' = \beta_x^{1/2} \frac{(B_y + \Delta B_y) l_s}{|B\rho|} = S\left(X^2 - \frac{\beta_y}{\beta_x} Y^2\right)\frac{|X|}{X} + \beta_x^{1/2} \frac{l_s}{|B\rho|} \Delta B_y \quad (19)$$

$$\Delta Y' = -2S\frac{\beta_y}{\beta_x}|X|Y.$$

By ignoring the influence of the vertical motion as it was done in Section 2.1, Eq. (19) can be rewritten as,

$$\Delta X = \Delta Y = \Delta Y' = 0$$

$$\Delta X' = S|X|X + \frac{2S\beta_x^{-1}}{\left(\frac{d^2 B_y}{dx^2}\right)_0} \Delta B_y. \quad (20)$$

Following the same derivation in Section 2.2, one can obtain the Kobayashi Hamiltonian including the field error,

$$X \geq 0 \qquad H = \varepsilon(X^2 + X'^2) - \frac{2}{3}SX^3 - \frac{2S\beta_x^{-1}}{\left(\frac{d^2 B_y}{dx^2}\right)_0}\Delta B_y X$$

$$X < 0 \qquad H = \varepsilon(X^2 + X'^2) + \frac{2}{3}SX^3 - \frac{2S\beta_x^{-1}}{\left(\frac{d^2 B_y}{dx^2}\right)_0}\Delta B_y X. \quad (21)$$

Fig. 16 shows the phase-space maps with and without the field error calculated from Eq. (21) at different tunes. One can see that when the tune is very close to a half-integer, the stable region becomes smaller and flatter with the field error; otherwise the influence of the field error can be ignored. That is to say, if the tune is close to a half-integer, more particles will be extracted with the field error and this might be considered as an advantage for extracting the beam core. This has also been confirmed by multi-particle simulations. Fig. 17 shows the number of extracted particles with turns for a tune close to the resonance, and one can see that the homogeneity of extracted beam is not influenced by the field error.

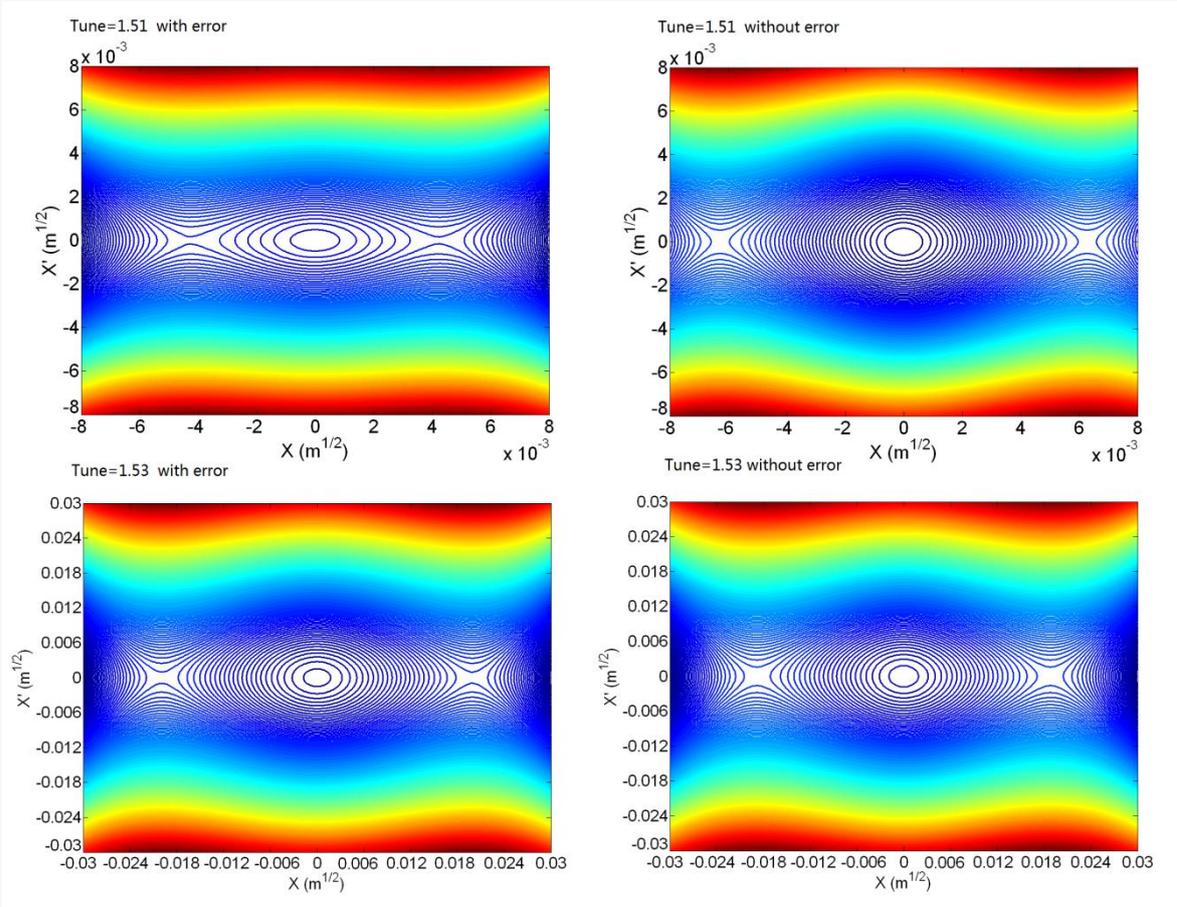

Fig. 16. The phase-space map with and without field error calculated from Eq. (21) ($S=10$ m$^{-1/2}$)

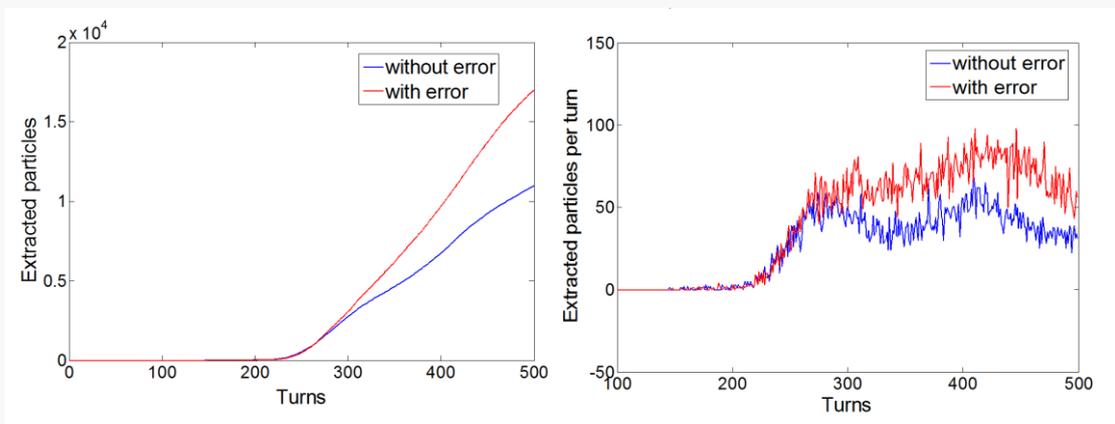

Fig. 17. Number of extracted particles with and without field error with turns (Tune = 1.505)

### 3.3.2. 2-D inherent field error

When the coupling between the horizontal and vertical phase planes is considered, one should give the 2-D field distribution of an anti-symmetric sextupole, as illustrated in Fig. 18. One can see that the vertical field component decreases along vertical direction with only a few per mille. Thus the number of extracted particles with 2-D field distribution has scarcely changed from the 1-D distribution, same for the cases of a standard sextupole or an anti-symmetric sextupole, which has been confirmed by multi-particle simulations. Therefore, it is expected that for an anti-symmetric sextupole, the combined function of the 2-D field distribution and inherent 2-D field error will be dominated by the 1-D error effect. In the simulations, the initial horizontal and vertical distributions are the same as before and the initial particle number is also 20000. Fig. 19 shows the distribution of 2-D inherent field error of the same anti-symmetric sextupole as above. Fig. 20 shows the number of extracted particles without and with 1-D/2-D field errors with respect to sextupole strength. It can be seen that when the stable region is very small, the 2-D field error slightly enhances the extraction than the 1-D field error.

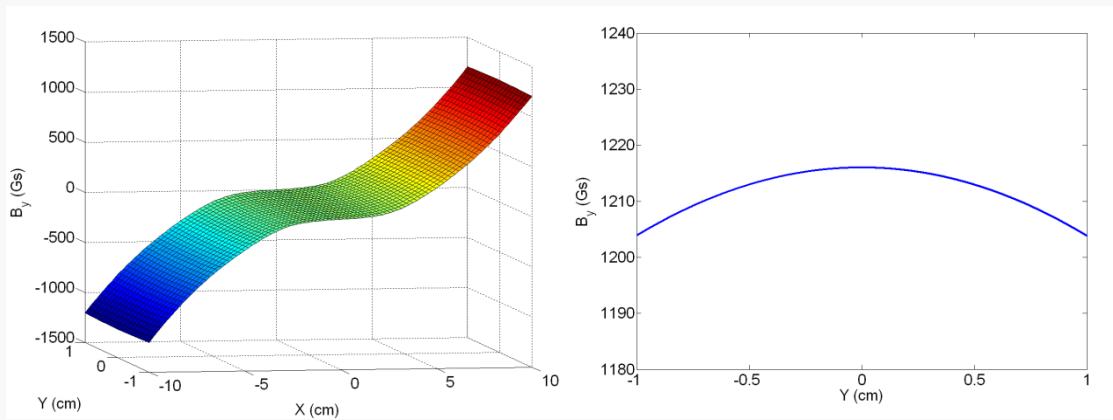

Fig. 18. 2-D field distribution of an ideal anti-symmetric sextupole (Left: 2-D field distribution; Right: field distribution at $X = 10$ cm)

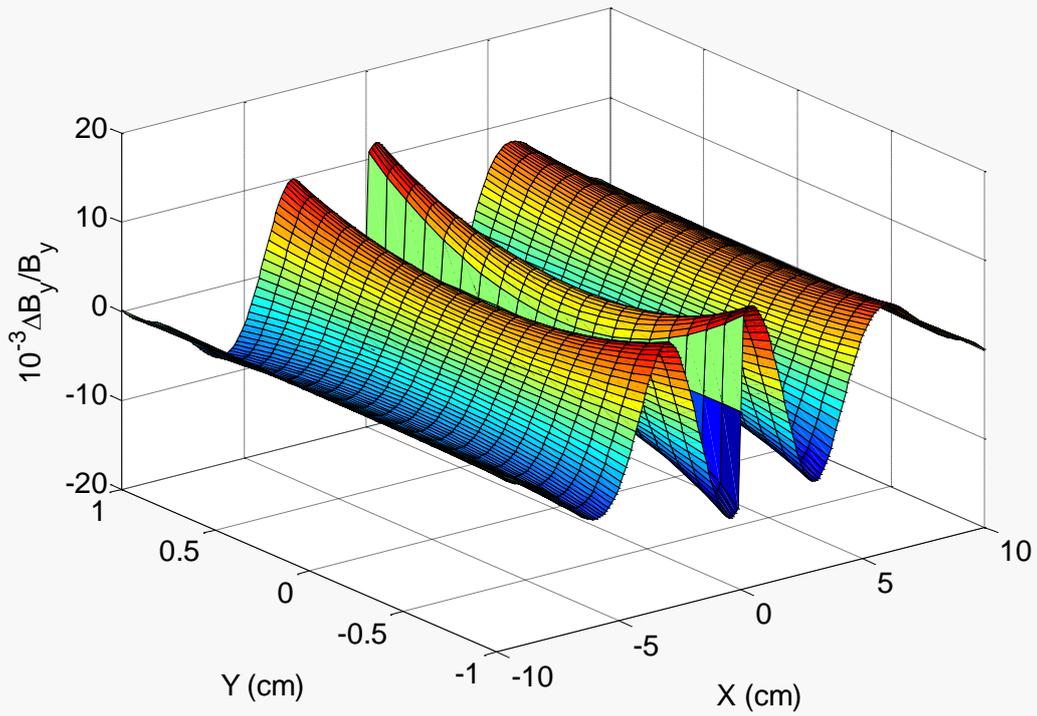

Fig. 19. Distribution of 2-D inherent field error of an anti-symmetric sextupole

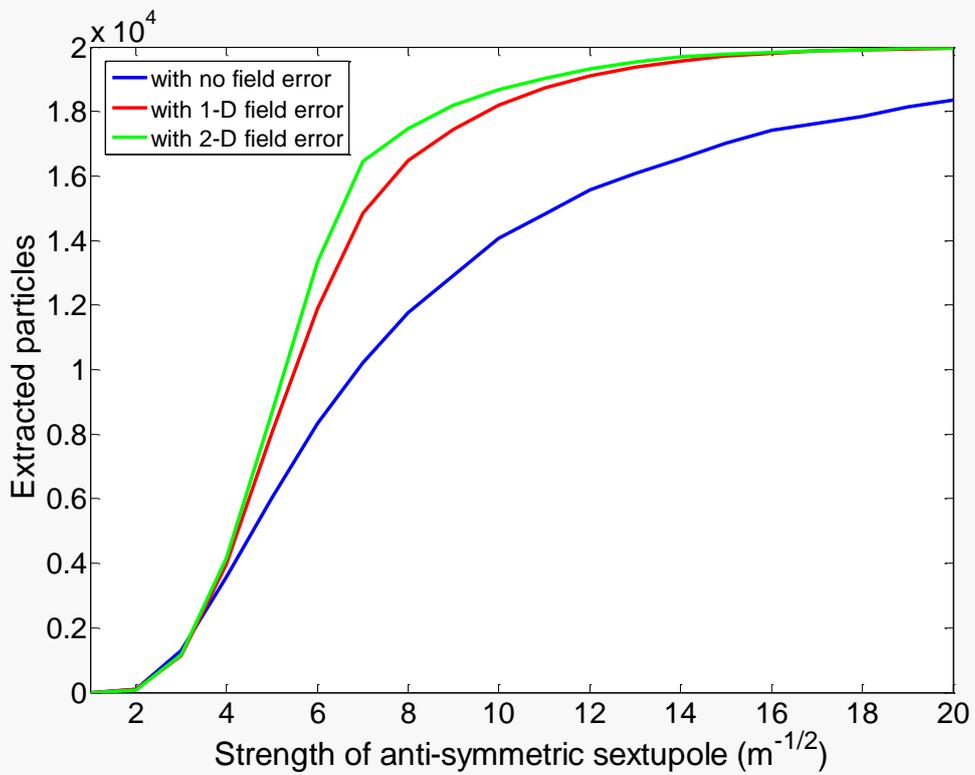

Fig. 20. Number of extracted particles with respect to sextupole strength (tune = 1.505)

*3.4. Half-integer stop band and resonance driven by quadrupole errors*

As we know, the half-integer resonance driven by quadrupole errors would happen if the tune is too close to a half-integer. This half-integer resonance can also be used to extract beam from a synchrotron [18-20], with help of octupoles a controllable stable region can be formed but the extraction is usually faster than the third-order resonant extraction [9, 21], and the quality of the extracted beam is usually less good. Therefore, the method is no longer adopted in modern synchrotrons. Here we just consider the resonance without octupole. Once the resonance starts all the particles will be driven outwards along a certain direction defined by the Courant-Snyder parameter [22] α and then extracted or lost in the ring, except it is stopped by moving away the tune from the resonance or reducing the driving quadrupole strength. Now the half-integer stop band is calculated to see whether normal quadrupole errors will cause the half-integer resonance when one applies an anti-symmetric sextupole to extract the beam. For the half-integer resonance driven by quadrupole errors, the stop band width can be calculated by the following formula [23]:

$$\Delta \nu_{1/2} = \left| \frac{1}{2\pi} \oint \beta(s) \Delta k(s) e^{-ip\varphi(s)} ds \right|, \quad (22)$$

where $\beta(s)$ is the betatron amplitude function, $\varphi(s)$ the betatron phase advance, $\Delta k(s)$ the field gradient error, and $p$ the harmonic number of the stopband integral. For discrete quadrupoles, and suppose only zero harmonic of the stopband integral is considered, Eq. (22) can be rewritten as,

$$\Delta \nu_{1/2} = \frac{1}{2\pi} \left| \sum_i \beta(s_i) \Delta k(s_i) \Delta L(s_i) \right|, \quad (23)$$

where $s_i$ and $\Delta L$ denote the quadrupole position and length, respectively. As for the example APTF, there are 4 pairs of quadrupoles symmetrically in the ring. For one pair of quadrupoles, using: $\beta(s_1) = 1.41$ m, $k(s_1) = 2.32$ m$^{-2}$; $\beta(s_2) = 6.88$ m, $k(s_2) = -1.81$ m$^{-2}$, $\Delta L(s_1,s_2) = 0.2$ m, one can estimate the bandwidth. In usual circumstances, the quadrupole gradient errors can be controlled lower than 0.1% (uniform distribution, maximum error), which corresponds to $\Delta k(s_1) = 2.32 \times 10^{-3}$ m$^{-2}$ and $\Delta k(s_2) = -1.81 \times 10^{-3}$ m$^{-2}$. Then one can obtain the maximum stop band width: $\Delta \nu_{1/2} = 0.0012$. In other words, a gradient error of 0.1% can cause resonance if the tune is in the range of 1.5±0.0006. Multi-particle simulations have also been carried out to show how the quadrupole error drives the half-integer resonance. Fig. 21 shows the relationship between the gradient error of quadrupole and the tune which defines the stopband, which is consistent to the above analytical result. As the working tune chosen for the resonant slow extraction by using an anti-symmetric sextupole is much more distant to the half-integer, it is safe to employ the method without needing to worry about the resonance driven by quadrupole errors.

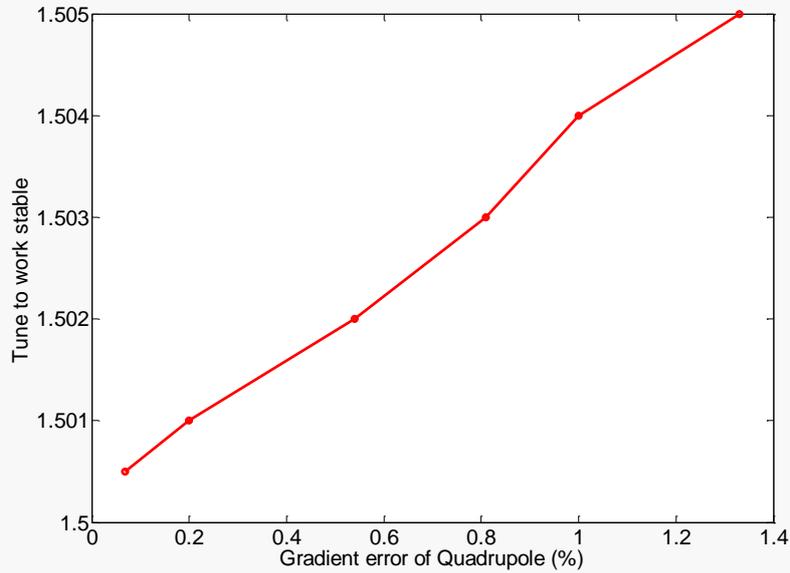

Fig. 21. Relationship between the gradient error of quadrupole and the tune which defines the stopband at a half-integer resonance

## 4. Conclusions and discussions

The studies show that the resonant slow extraction in synchrotrons by using anti-symmetric sextupole fields is feasible and has some important advantages as compared to the usual third-order resonance extraction by standard sextupoles, e.g. more distant tune to the resonance which means less sensitivity to the tune variation, more flexibility to place septum magnets, and faster response in resonance excitation. Multi-particle simulations and the phase-space maps calculated by the Kobayashi Hamiltonian agree each other and support the concept. An empirical formula for the stable region of the half-integer resonance driven by an anti-symmetric sextupole has been extracted from the Kobayashi Hamiltonian and verified by simulation results. From the formula, the stable region is proportional to $(\delta Q/S)^2$, same as in the case of the third-order resonance by a standard sextupole. However, the coefficient is only about 1/14 of the one in the third-order resonance. This gives the advantages for the former, as the tune can be placed more distant from the resonance and the resonance is less sensitive to the tune variation by power supply ripples. Both 1-D and 2-D inherent field errors of anti-symmetric sextupoles have been analyzed and it is found that they do not hinder their application in extracting beam from a synchrotron. The study also shows that the half-integer resonance is overwhelm dominated by the anti-symmetric sextupole and almost not affected by usual quadrupole gradient errors, as long as the tune is kept away from the resonance by at least 0.001. The RFKO method is still recommended to assist the resonant slow extraction by an anti-symmetric sextuple, just same as in the case of the usual third-order resonance. One can also think that with replacing octupole by anti-symmetric sextupole the half-integer resonant extraction method retakes the advantage over the third-order resonant extraction method.

Although the feasibility of using an anti-symmetric sextupole in the slow extraction from a synchrotron has been demonstrated in this article, further studies are needed to make the method fully applicable in real machines. Other applications with such special half-integer resonance, such as halo collimation in large proton-proton colliders, are also needed to be exploited.

**Acknowledgments**

The authors would like to thank Alex W. Chao of SLAC for very useful discussions and comments, and also the other students in the group for discussions. The study is supported by National Natural Science Foundation of China (Project 11235012)

**References**

[1] M. Gyr, W. Bartmann et al., Resonant third-integer extraction from the PS2, Proceedings of PAC09, Vancouver, BC, Canada, 2009, p. 1593

[2] G. Feldbauer, M. Benedikt, U. Dorda, "Simulations of Various Driving Mechisms for the 3rd Order Resonant Extraction from the MedAustron Medical Synchrotron", Proceedings of IPAC2011, San Sebastián, Spain, P3481

[3]M. Tomizawa et.al., Slow beam extraction at TARN II, NIM-A, 326-3(1993), p. 399-406

[4] M. Tomizawa et al., Status and upgrade plan of slow extraction from the J-PARC main ring, Proceedings of IPAC10, Kyoto, Japan, 2010, p. 3912

[5] G. Zhen, J.Y. Tang et.al., A novel structure of multipole field magnets and their applications in

uniformizing beam spot at target, NIM-A, 691(2012), p. 97-108

[6] Y. Zou, J.Y. Tang et. al., Resonant Slow Extraction in Synchrotrons by Using Anti-symmetric Sextupole Fields, Proceedings of IPAC14, Dresden, Germany (2014), p. 827

[7] Future Circular Collider Study Hadron Collider Parameters, CERN, FCC-ACC-SPC-0001, 2014

[8] Pre-Conceptual Design Report for CEPC-SPPC, IHEP, 2015

[9] Accelerator Complex Study Group, Proton-Ion Medical Machine Study (PIMMS) Part I, CERN/PS-99-010(DI), (2000)

[10] Y. Kobayashi, H. Takahashi, Improvement of the emittance in the resonant ejection, Proc. VIth Int. Conf. on High Energy Accelerators, Cambridge, Massachusetts, (1967), p347-51.

[11] K. Noda et al., Advanced RF-KO slow-extraction method for the reduction of spill ripple, NIM A 492, (2002) p. 253

[12] T. Furukawa et. al., Global spill control in RF-knockout slow extraction, NIM-A, 522-3(2004), p. 196-204

[13] M. Kirk et al., SIS-18 RF Knock-Out optimisation studies, Proceedings of IPAC2013, Shanghai, China, (2013) p. 297

[14] V.P. Nagaslaev, J.F. Amundson, J.A. Johnstone, C.S. Park, S.J. Werkema, "Third Integer Resonance Slow Extraction Using RFKO at High Space Charge", Proceedings of IPAC2011, San Sebastián, Spain, P3559

[15] A. Miyamoto et al., Study of slow beam extraction through the third order resonance with transverse phase space manipulation by a mono-frequency RFKO, Proceedings of PAC2005, Knoxville, Tennessee, (2005) p. 1892

[16] A. Noda et al., Slow beam extraction at KSR with combination of third order resonance and RFKO, Proceedings of PAC1999, New York, USA, (1999) p. 1294

[17] FANG Shou-Xian, GUAN Xia-Ling, TANG Jing-Yu, CHEN Yuan, DENG Chang-Dong, DONG Hai-Yi, FU Shi-Nian, JIAO Yi, SHU Hang, OUYANG Hua-Fu, QIU Jing, SHI Cai-Tu, SUN Hong, WEI Jie, YANG Mei, ZHANG Jing, ATPF - a Dedicated Proton Therapy Facility, Chinese Physics C, 2010, Vol.34 (03): 383-388

[18] H. T. Edwards et al., Half-integer resonance extraction for the NAL accelerator, Proceedings of PAC1973, San Francisco, CA (1973) p. 424

[19] L. Michelotti, Classification of half-integer resonance dynamics, Proceedings of PAC1985,

Vancouver, BC, Canada, (1985) p. 2258

[20] P. I. Gladkikh et al., A slow half-integer resonance beam extraction in a pulse stretcher ring PSR-2000, Proceedings of EPAC1990, Nice, France, (1990) p. 1691

[21] R. Cappi et al., Multiturn extraction: performance analysis of old and new approaches, NIM, A519 (2004) 442-452

[22] E.D. Courant and H.S. Snyder, Theory of the Alternating-Gradient Synchrotron, Annals of Physics: 3, (1958) 1-48

[23] Helmut Wiedemann, Particle Accelerator Physics, 3th edition, Springer Berlin Heidelberg, New York, (2007) p. 427-432